\documentclass{article}

\usepackage[preprint]{neurips_2025}

\usepackage[utf8]{inputenc} 
\usepackage[T1]{fontenc}
\usepackage{amsmath}
\usepackage{hyperref}
\usepackage{url}
\usepackage{booktabs}
\usepackage{amsfonts} 
\usepackage{nicefrac} 
\usepackage{microtype} 
\usepackage{xcolor} 
\usepackage{colortbl} 
\usepackage{hhline}
\usepackage{graphicx}
\usepackage{fontawesome}
\usepackage{placeins}
\usepackage{svg}
\usepackage{tcolorbox}
\usepackage{soul}
\usepackage{subcaption}
\usepackage{makecell}
\usepackage{booktabs}
\usepackage{array}

\newcommand{\benname}{FinAR-Bench}

\title{Towards Competent AI for Fundamental Analysis in Finance: A Benchmark Dataset and Evaluation}

\author{
Zonghan Wu\textsuperscript{1}, Congyuan Zou\textsuperscript{1}, Junlin Wang\textsuperscript{1}, Chenhan Wang\textsuperscript{2},
Hanjing Yang\textsuperscript{3}, Yilei Shao\textsuperscript{1}\thanks{Corresponding author}\\
\textsuperscript{1}Shanghai AI Finance School, East China Normal University\\
\textsuperscript{2}OpenBayes.com,
\textsuperscript{3}Tsinghua University\\
\texttt{\{zhwu, jlwang\}@sem.ecnu.edu.cn, 10224800458@stu.ecnu.edu.cn}\\
\texttt{gabriel@openbayes.com, yang-hj24@mails.tsinghua.edu.cn}\\
\texttt{yileishao@sem.ecnu.edu.cn}
}

\begin{document}

\maketitle

\begin{abstract}

Generative AI, particularly large language models (LLMs), is beginning to transform the financial industry by automating tasks and helping to make sense of complex financial information. One especially promising use case is the automatic creation of fundamental analysis reports, which are essential for making informed investment decisions, evaluating credit risks, guiding corporate mergers, etc. While LLMs attempt to generate these reports from a single prompt, the risks of inaccuracy are significant. Poor analysis can lead to misguided investments, regulatory issues, and loss of trust. Existing financial benchmarks mainly evaluate how well LLMs answer financial questions but do not reflect performance in real-world tasks like generating financial analysis reports. In this paper, we propose FinAR-Bench, a solid benchmark dataset focusing on financial statement analysis, a core competence of fundamental analysis. To make the evaluation more precise and reliable, we break this task into three measurable steps: extracting key information, calculating financial indicators, and applying logical reasoning. This structured approach allows us to objectively assess how well LLMs perform each step of the process. Our findings offer a clear understanding of LLMs current strengths and limitations in fundamental analysis and provide a more practical way to benchmark their performance in real-world financial settings.

\end{abstract}

\section{Introduction}
\label{sec:intro}

The growing capabilities of generative AI are beginning to reshape the financial sector \cite{nie2024survey}, where LLMs offer promising opportunities to answer financial questions \cite{wu2023bloomberggpt}, enhance decision-making processes \cite{yu2024fincon}, and generate insights from multi-modal financial data \cite{bhatia-etal-2024-fintral}.  Particularly, the automatic generation of fundamental analysis reports represents a high-value application within this area. Fundamental analysis is a method of assessing companies by examining economic environments, financial statements, market positions, and other qualitative and quantitative factors. It is a critical tool for making long-term investments, credit assessments, corporate merger decisions, etc.

With the emergence of LLMs, one can attempt to generate a fundamental analysis report of a publicly listed company with a single prompt. However, the stakes are extremely high. Inaccurate financial analysis can lead to misguided investment decisions, regulatory compliance issues, and erosion of stakeholder trust. This creates a tension between the promising capabilities of LLMs and the stringent requirements for precision, reliability, and transparency in financial contexts.  As organizations move from experimental to production applications of these technologies, there is an urgent need to establish reliable benchmarks that can properly assess the capabilities of LLMs in performing financial fundamental analysis.

Existing finance-focused benchmarks primarily assess LLMs on their ability to answer expert-level questions, covering areas such as financial natural language understanding \cite{shah_when_2022,lu_bbt-fin_2023,xie_pixiu_2023}, numerical reasoning \cite{zhu_tat-qa_2021, chen_finqa_2021}, and professional certification tests \cite{xu_superclue-fin_2024,nie_cfinbench_2024}. Efforts to increase the complexity of these evaluations have included enlarging context lengths \cite{chen_fintextqa_2024}, introducing intricate table structures \cite{zhao_multihiertt_2022}, formulating knowledge-intensive mathematical problems \cite{zhao_financemath_2024}, and covering comprehensive financial qualification tests \cite{nie_cfinbench_2024}.  These benchmarks emphasize question diversity to evaluate the generalization capabilities of LLMs within the financial domain.  However, recent trends are shifting from question solvers to task assistants. Current financial benchmarks are insufficient in evaluating LLMs' performance within such task-specific settings such as fundamental analysis.

In this paper, we aim to evaluate the core capabilities of LLMs in performing typical fundamental analysis tasks.  Fundamental analysis usually consists of several main sections: economic conditions, industry analysis,  business analysis, and financial statement analysis. We focus on  financial statement analysis as it is the most important part of fundamental analysis.  If an LLM is unable to perform well in financial statement analysis, its effectiveness in broader fundamental analysis tasks is likely limited.  Financial statement analysis is a diagnosis of a company's financial health and performance based on its balance sheet, income statement, and cashflow statement, which usually appears in its periodical reports such as the annual report. Directly assessing the quality of a financial statement analysis generated by LLMs can be challenging, as the generated content is probabilistic and lacks definitive ground truth comparisons. To enable a more rigorous assessment, we decompose financial statement analysis into three subtasks: information extraction, indicator calculation, and logical reasoning. Each subtask yields structured intermediate outputs that can be objectively verified against ground truth data. Using these verifiable steps, we derive an implicit estimate of the capability of LLMs in conducting financial statement analysis. The main contributions of this paper are summarized as follows,

\begin{itemize}
    \item We propose \textbf{\benname{}} (\textbf{Fin}ancial \textbf{A}nalysis and \textbf{R}easoning \textbf{Bench}mark), a task-oriented LLM benchmark dataset in 
    financial fundamental analysis with its first set of benchmarks targeting at evaluating LLMs' capabilities in financial statement analysis.
    \item We conduct a solid evaluation to assess the capabilities of current LLMs in financial statement analysis. Experimental results indicate that LLMs perform well in information extraction, struggle with indicator computation, and exhibit promising potential in logical reasoning.
    \item We release the source code and dataset for the research community, to encourage future open work to utilize our dataset for self-verification. Our code and data are publicly accessible at \href{https://github.com/SAIFS-AIHub/FinAR-Bench}{https://github.com/SAIFS-AIHub/FinAR-Bench}.

\end{itemize}

\section{Related Works}
\label{relatedworks}

Related works on financial evaluation benchmarks can be categorized into three groups: financial language understanding, financial knowledge and application, and financial numerical reasoning.

\textbf{Financial Language Understanding} involves the processing and comprehension of text within financial contexts, which often contains domain-specific terminology and complex concepts. The FLUE benchmark \cite{shah_when_2022}, introduced alongside the FLANG model, contains five distinct NLP tasks: financial sentiment analysis \cite{malo2014good}, news headline classification \cite{sinha2021impact}, named entity recognition \cite{alvarado2015domain}, structure boundary detection, and question answering. 
The BBT-CFLEB benchmark includes six financial NLP tasks covering both understanding and generation tasks \cite{lu_bbt-fin_2023}.
The Flare benchmark encompasses eight critical financial tasks, including six financial NLP tasks and two financial prediction tasks, evaluated across 15 different datasets \cite{xie_pixiu_2023}.

\textbf{Financial Knowledge \& Application} benchmarks mostly involve financial knowledge qualification tests and diverse financial applications. FinTextQA is developed to focus specifically on long-form question answering in finance \cite{chen_fintextqa_2024}. It stands out for its comprehensive coverage of complex financial question systems, including queries on financial regulations and policies that often require detailed explanations rather than simple numerical answers. 
FinEval contains questions carefully categorized into four key areas: financial academic knowledge, financial industry knowledge, financial security knowledge, and financial agent \cite{guo_fineval_2024}.
SuperClue-Fin assesses models across six financial application domains and twenty-five specialized tasks, covering both theoretical knowledge and practical knowledge applications such as compliance, risk management, and investment analysis \cite{xu_superclue-fin_2024}.
CFinBench establishes a four-dimensional evaluation system mirroring the knowledge progression of Chinese financial professionals \cite{nie_cfinbench_2024}. The benchmark comprises financial subject, financial qualification, financial practice, and financial law. FLAME introduces complementary evaluation dimensions through FLAME-Cer and FLAME-Sce \cite{guo_flame_2025}, where FLAME-Cer is a Certification-focused assessment across 14 financial qualifications and FLAME-Sce is a practical scenario evaluation covering 10 core financial business scenarios. FinBen encloses extensive financial datasets across seven categories, information extraction, textual analysis, question answering, text generation, risk management, forecasting, and decision-making \cite{xie2024finben}.  The text generation task in FinBen mainly focuses on text summarization.  

\textbf{Financial Numerical Reasoning} focuses on various computations of numerical data within financial contexts.
TAT-QA represents one of the first financial question-answering datasets over hybrid data formats \cite{zhu_tat-qa_2021}. Its primary contribution lies in its focus on hybrid contexts, where answering questions requires integrating information from both tabular data and associated textual paragraphs. 
FinQA addresses the challenge of deep numerical reasoning over financial data \cite{chen_finqa_2021}. The dataset is created by financial experts and focuses on complex multi-step calculations required to answer questions about financial reports. While TAT-QA and FinQA typically include only a single flat table in each document, MultiHiertt incorporates multiple hierarchical tables alongside textual content, more accurately reflecting the complexity of real-world financial documents \cite{zhao_multihiertt_2022}.
Finance-Math evaluates LLMs' capabilities in solving knowledge-intensive math reasoning problems \cite{zhao_financemath_2024}. It provides expert-annotated solution references in Python program format, ensuring a high-quality standard for evaluation.  DocMath-Eval focuses on the numerical reasoning capabilities of LLMs within financial document contexts \cite{zhao_docmath-eval_2024}. The benchmark comprises four evaluation sets with varying levels of difficulty in both numerical reasoning and document understanding.  BizBench proposes an eight-task evaluation pyramid focusing on programmatic financial problem-solving including program synthesis, quantity extraction, and domain knowledge \cite{krumdick_bizbench_2024}.
FinDVer evaluates claim verification capabilities of LLMs in the context of understanding and analyzing long, hybrid-content financial documents \cite{zhao_findver_2024}.  Given hybrid-content financial documents, LLMs are tasked with classifying financial claims as "entailed" and "refuted".

\section{\benname{}}
\label{\benname{}}
\subsection{Financial Statement Data}
Financial statement data of a company contains three main tables, income statement, balance sheet, and cash flow statement. They together show a company's profitability, financial position, and cash movements.
We collect financial statement data of one hundred companies in the fiscal year 2023 from the Shanghai Stock Exchange (SSE) website\footnote{https://www.sse.com.cn/}. The SSE provides corporate financial statements in two formats: XBRL and PDF. This dual availability allows us to benchmark LLMs using both textual and file-based data. 

The XBRL form data is a standardized format used for exchanging and communicating financial data electronically. As XBRL data is a structured format, it allows us to generate ground truth labels without human labeling.  On the contrary, PDF-formed financial reports are unstructured and vary in style and layout across companies, often spanning hundreds of pages. To maintain a manageable cost for benchmark evaluation, we selectively extract only the pages containing balance sheet table, income statement table, and cash flow statement table from these reports.

\subsection{Task 1: Information Extraction}
Information extraction is a fundamental yet labor-intensive task for financial analysts. Since it forms the foundation of financial statement analysis, it demands a high level of precision. With this in mind, we design an information extraction task that requires an LLM to extract multiple financial items from financial statement data. The goal is to evaluate how reliably an LLM can read financial documents and accurately transform the information into a structured format. To achieve this, we design clear and precise requirements for an LLM. The task prompt consists of three parts, the task description,  the task requirement, and financial statement data. A demonstration is given in the following: 
\begin{quote}
Extract the company's [revenue, cost of revenue, net income, cash and cash equivalents, accounts receivable, accounts payable, total assets, total liabilities, and net cash flow from operating activities, ...] in 2022 and 2023 from the attached data. Output the results in a markdown-formatted table. Use 'Item', '2022', and '2023' as the column headers. [Finanical statement data].
\end{quote}

\subsection{Task 2: Indicator Computation}
To gain critical insights into a company's operating status, it is essential to calculate various ratios, proportions, and year-over-year changes. When an LLM is tasked with conducting an analysis of a company's financial statements, this process typically involves calculating and reporting a range of key financial indicators. 

Considering it is challenging to evaluate AI-generated contents in unconstrained form, we design the indicator computation task by directing an LLM to produce a series of specific indicators in a controlled and standardized manner. This simplified task could serve as a foundational step toward establishing trust in AI-generated financial analysis outputs.
Similar to Task 1, the prompt of Task 2 takes the following form,

\begin{quote}
    Calculate the company's [return on equity, return on assets, gross margin, net profit margin, revenue growth rate, net profit growth rate, debt to assets, debt to equity, equity to assets, current ratio, quick ratio, inventory turnover, receivables turnover, ...] in  2023 given the attached data. Output the results in a markdown-formatted table. Use 'Item', '2023' as the column headers. Express the result as a decimal, rounded to four decimal places. [Finanical statement data].
\end{quote}

\subsection{Task 3: Logical Reasoning}

A fundamental principle of analysis is that it should begin with the careful observation and identification of facts, followed by the interpretation of those facts to draw conclusions or gain a deeper understanding. In financial statement analysis, the observation about a company usually consists of three elements: comparison with the company's performance in the previous time period, comparison between different items within the same category, and comparison with industry averages. For example, an increase in a company's return on equity can be interpreted as an enhancement of its profitability. Building on this insight, we introduce a logical reasoning task that first instructs the LLM to observe facts under clearly defined judging conditions and then to reason over the satisfied conditions to draw meaningful interpretations. The prompt design is illustrated below:

\begin{quote}
Given the judging condition and the company's financial data, complete the following tasks:
I. Assess if the company's financial status in 2023 meets each of the specified conditions, and present the results in a markdown-formatted table with two columns: No. and Condition Met. 
II. Based on the results from Step I, conduct an in-depth and comprehensive analysis and interpretation of the conditions that are met.
The judging conditions are given as follows,
1. Return on equity increases;
2. Return on total assets increases;
3. Gross profit margin increases;
4. Net profit margin increases;
5. Revenue growth rate > 0;
6. Net profit growth rate > 0;
7. Current ratio increases;
8. Quick ratio increases;
9. Debt-to-asset ratio increases;
...
[Finanical statement data]
\end{quote}

\section{Evaluation Approach}
\subsection{Table Assessment}

To accurately evaluate the capabilities of LLMs in financial statements analysis, we design a structured table evaluation protocol. Specifically, we prompt LLMs to produce outputs in Markdown table format, explicitly specifying column headers. The resulting Markdown tables are normalized to a standardized format, enabling systematic comparison against the ground truth tables. In this study, we utilize the RMS metric for evaluation \cite{liu-etal-2023-deplot}. Originally developed for chart-to-table research, RMS simultaneously accounts for key-value structural alignment and accuracy, making it well-suited for assessing our benchmark.

\paragraph{RMS Metric Computation.}

The RMS computation follows several key steps:

\begin{enumerate}
    \item \textbf{Data Point Extraction:}  
    Each table is parsed into a set of data points, where each point consists of a \textit{row header}, a \textit{column header}, and a \textit{numerical value}. Row and column headers are concatenated to form a unique key (e.g., "Sales Expense 2022").

    \item \textbf{Textual Distance Calculation:}  
    For each predicted-target key pair, compute the Normalized Levenshtein Distance:
    \begin{equation}
    \text{NL}(pr \Vert pc,\ tr \Vert tc) = 
    \frac{\operatorname{edit\_distance}(pr \Vert pc,\ tr \Vert tc)}{
    \max\big(\operatorname{len}(pr \Vert pc),\ \operatorname{len}(tr \Vert tc)\big)}.
    \end{equation}
    
    The subsequent cost matrix used in the assignment step is computed as:
    \begin{equation}
    \text{Cost}(pr, pc, tr, tc) =
    \begin{cases}
    \text{NL}(pr \Vert pc,\ tr \Vert tc), & \text{if } \text{NL}(pr \Vert pc,\ tr \Vert tc) \leq \tau \\
    1, & \text{otherwise}
    \end{cases}
    \end{equation}

    where $(pr, pc)$ and $(tr, tc)$ denote the predicted and target concatenated headers, respectively.

    \item \textbf{Optimal Assignment via Hungarian Algorithm:}  
    Using the textual distance cost matrix, apply the Hungarian algorithm \cite{kuhn1955hungarian} to determine the optimal one-to-one assignment between predicted and target data points, minimizing the total matching cost.

    \item \textbf{Numerical Error Calculation:}  
    For each assigned pair, compute the relative error: 

    \begin{equation}
    D_{\theta}(p, t) = 
    \begin{cases}
    \frac{||p - t||}{||t||}, & \text{if } \frac{||p - t||}{||t||} \leq \theta \\
    1, & \text{otherwise}
    \end{cases}
    \end{equation}

    where $p$ and $t$ are the predicted and the ground truth numerical values.

    \item \textbf{Final RMS Precision and Recall:}  
    
    Unlike the original RMS formulation which combines textual distance and numerical deviation, we remove the textual distance component from the final score to better focus on numerical accuracy. Specifically, we define:

    \begin{equation}
    D_{\tau, \theta}(p, t) = 
    \begin{cases}
    D_{\theta}(p, t), & \text{if } \text{NL}(pr || pc, tr || tc) \leq \tau \\
    1, & \text{otherwise}
    \end{cases}
    \end{equation}

    The revised RMS Precision and RMS Recall are computed as:
    \begin{equation}
    \text{RMS}_{\text{Precision}} = 1 - \frac{\sum_{i=1}^{N} \sum_{j=1}^{M} X_{ij} D_{\tau, \theta}(p_i, t_j)}{N},
    \end{equation}
    \begin{equation}
    \text{RMS}_{\text{Recall}} = 1 - \frac{\sum_{i=1}^{N} \sum_{j=1}^{M} X_{ij} D_{\tau, \theta}(p_i, t_j)}{M},
    \end{equation}
    where $X_{ij}$ denotes the assignment matrix obtained from the Hungarian algorithm, $N$ is the number of predicted data points, and $M$ is the number of ground-truth data points.
\end{enumerate}

\subsection{Reasoning Assessment}
In the logical reasoning task, apart from evaluating the correctness of table outputs, we must also assess the quality of the accompanying analysis. Scoring this analysis on a scale from 0 to 100 is inherently difficult for both humans and machines, as it requires fine-grained, subjective judgment and consistent criteria across diverse cases. However, making relative comparisons between two model outputs given the same prompt is significantly more manageable. To leverage this, we set up a tournament-style evaluation using LLM-as-a-judge method \cite{zheng2023judging}, enabling systematic pairwise comparisons to identify the best reasoning model for financial statement analysis.
\paragraph{Tournament Ranking}
The tournament is structured as a round-robin competition, where each candidate is matched against every other candidate in a series of pairwise comparisons. In each match, both candidates are given the same prompt, and their responses are evaluated by a LLM judge. The judge determines which response is better. The scoring system is simple: the winner of each pairwise match receives 1 point, the loser receives 0 points, and in the case of a tie, both candidates receive 0.5 points. After all matches are completed, the total scores for each candidate are calculated, and candidates are ranked based on their overall points.
\paragraph{LLM-as-a-Judge}
The LLM judge assumes the role of a financial expert and evaluator. The task is to compare two financial statement analyses generated by two candidate models and determine which one is superior. Each comparison involves reviewing both analyses based on three key criteria: 

\begin{itemize}
    \item \textbf{Consistency}: Evaluate whether each report is internally coherent and logically sound. This includes checking the alignment between numerical data, identified trends, and stated conclusions, as well as the adherence to standard financial reasoning practices.
    \item \textbf{Depth of Analysis}: Assess the extent to which each report transcends basic fact listing to construct an integrated financial diagnosis. A strong analysis should meaningfully connect multiple financial indicators and reveal the relationships among key metrics.
    \item \textbf{Financial Insight}: Determine whether the report offers thoughtful and non-obvious interpretations. Reports with high insight demonstrate financial expertise, critical thinking, and the ability to explain underlying causes, implications, or strategic considerations behind the numbers.
\end{itemize}
\section{Experiments}
\subsection{Experimental Setup}

\paragraph{Dataset}
We curate a dataset containing one hundred companies listed on the Shanghai Stock Exchange for the fiscal year 2023. We prepare the financial statement data of each company in two forms, the textual table form converted from its XBRL data and the raw PDF form extracted from its annual report. The dataset is divided into a development set and a test set with a ratio of 1:9. 
\paragraph{Baselines}
We select 14 LLMs for evaluation,  categorized by their parameter sizes into three groups: Large (>100B), Medium (>10B), and Small (<10B). These models include: \textbf{ChatGPT Series}: GPT-4o, GPT-o1. \textbf{DeepSeek Series}: DeepSeek-v3, DeepSeek-r1, DeepSeek-r1-distill-qwen-32b, DeepSeek-r1-distill-qwen-14b, DeepSeek-r1-distill-llama-8b. \textbf{Llama Series}: Llama-3.1-405b-instruct, Llama-3.1-8b-instruct. \textbf{Mistral Series}: Mistral-7b-instruct-v0.3, Mixtral-8*22b-instruct-v0.1, Mixtral-8*7b-instruct-v0.1. \textbf{Qwen Series}: Qwq-32b, Qwen2.5-7b-instruct.

\textbf{Experiment Settings} For open-sourced LLMs, we evaluate them through NVIDIA NIM API. The models are used with default parameter settings, and \texttt{maxtokens} is set to the maximum value to prevent output truncation during inference. Since LLMs can not handle PDF input directly, we use PyMuPDF \footnote{https://pymupdf.readthedocs.io/en/latest/} as the PDF extractor, which outperforms other alternatives according to our experiment, refer to Appendix \ref{sec: pdf}.  During the evaluation, we set the text matching threshold $\tau=1$,  ensuring that all key pairs are considered eligible during the assignment phase.
We set the numerical error threshold $\theta=0$, reflecting high precision requirements in finance.

\subsection{Main Results}

Table \ref{tab:text-evaluation} shows the main results of baseline LLMs on the test set of \benname{} under the text format. Results under the PDF format are provided in Appendix \ref{sec: pdf}. In the information extraction task,  large-sized LLMs achieve near-perfect scores, while medium-sized models remain competent. In contrast, small-sized LLMs exhibit significant performance degradation, primarily due to their limited capacity to process and generate long texts. In the indicator computation task, all models regardless of size perform poorly, reflecting a general deficiency in precise numerical computation.  To further investigate LLMs' capability in approximate numerical computation, we perform additional evaluation by varying error tolerance threshold (refer to Table \ref{tab: error}). Performance on the logic reasoning task surpasses that of the indicator computation task, largely due to a greater tolerance for numerical imprecision. This finding aligns with patterns observed in LLM-generated financial analysis reports, where the logical reasoning often appears sound even when the quantitative details are inaccurate. These results suggest that LLMs may still offer valuable insights in financial contexts, despite their limitations in exact calculation. Moreover, model performance is consistently lower in the PDF setting compared to the text setting as the variability of PDF layout adds complexity.

\begin{table*}
\centering
\footnotesize
\caption{Precision and recall of all models under text input.}
\label{tab:text-evaluation}
\resizebox{0.98\textwidth}{!}{
\begin{tabular}{ll
cc  
cc  
cc  
}
\toprule
\textbf{Size} & \textbf{Model} 
& \multicolumn{2}{c}{\textbf{Information Extraction}} 
& \multicolumn{2}{c}{\textbf{Indicator Computation}} 
& \multicolumn{2}{c}{\textbf{Logic Reasoning}} \\
\cmidrule(lr){3-4} \cmidrule(lr){5-6} \cmidrule(lr){7-8}
& 
& P & R 
& P & R 
& P & R \\
\midrule
\textbf{L} & GPT-4o & 98.39 & 98.39 & 34.38 & 34.38 & 63.23 & 63.23 \\
& GPT-o1 & \textbf{100.00} & \textbf{100.00} & 34.76 & 34.76 & \textbf{86.20} & \textbf{86.20} \\
& DeepSeek-v3 & 99.98 & \underline{99.98} & \underline{38.26} & \underline{38.26} & 63.11 & 63.11 \\
& DeepSeek-r1 & \textbf{100.00} & 99.93 & \textbf{49.31} & \textbf{49.31} & \underline{83.83} & \underline{76.28} \\
& Llama-3.1-405b-instruct & 99.03 & 99.03 & 20.83 & 20.80 & 64.90 & 63.84 \\
& Mixtral-8*22b-instruct-v0.1 & 97.34 & 98.19 & 21.96 & 21.94 & 48.05 & 47.94 \\
\midrule
\textbf{M} & DeepSeek-r1-distill-qwen-32b & 99.01 & 98.77 & 26.48 & 26.42 & 76.40 & 71.27 \\
& Qwq-32b & 98.96 & 98.85 & 35.01 & 34.97 & 77.50 & 75.91 \\
& Mixtral-8*7b-instruct-v0.1 & 86.48 & 84.31 & 11.01 & 10.87 & 47.81 & 39.76 \\
& DeepSeek-r1-distill-qwen-14b & 95.33 & 94.51 & 17.67 & 17.47 & 73.00 & 71.74 \\
\midrule
\textbf{S} & DeepSeek-r1-distill-llama-8b & 63.64 & 70.57 & 6.91 & 6.70 & 58.09 & 50.69 \\
& Llama-3.1-8b-instruct & 37.81 & 67.22 & 8.46 & 8.40 & 50.19 & 50.19 \\
& Qwen2.5-7b-instruct & 65.66 & 79.79 & 8.70 & 8.82 & 51.92 & 51.89 \\
& Mistral-7b-instruct-v0.3 & 64.45 & 79.13 & 2.37 & 2.40 & -- & -- \\
\bottomrule
\end{tabular}
}
\end{table*}

\paragraph{Reasoning analysis ranking}
We conduct a tournament-style evaluation of reasoning outputs using \texttt{Doubao-1.5-thinking-pro}, \texttt{Qwen-Max} and \texttt{Gemini-2.5-Pro} as the judges. Only large-sized models with recall scores above 60\% are included in the competition. To avoid order bias, each pair of models competes in two matches: one where model A is evaluated against model B, and another with the order reversed (model B v.s. model A).
This setup yields a total of 1800 competitions among 5 LLMs per judge. As shown in Table~\ref{tab:tournament-reasoning}, \texttt{GPT-o1} and \texttt{DeepSeek-r1} leads the ranking with 634.3 and 624.5 wins, respectively. \texttt{DeepSeek-r1} and \texttt{GPT-4o} shows moderate performance, while \texttt{Llama-3.1-405b-instruct} lags far behind. 

\begin{table}
\centering
\scriptsize
{
\caption{Scores from three LLM judges and summary statistics. Abbreviations: D = Doubao-1.5-pro, Q = Qwen-Max, G = Gemini-2.5-Pro, Avg = Average score, Tok = Avg. generated tokens, Rk = Rank.}
\label{tab:tournament-reasoning}
\resizebox{0.7\textwidth}{!}{
\begin{tabular}{lcccccc}
\toprule
\textbf{Model} & \textbf{D} & \textbf{Q} & \textbf{G} & \textbf{Tok} & \textbf{Avg / Rk} \\
\midrule
GPT-o1         & \textbf{649} & \textbf{637} & 617 & 9681.9 & \textbf{634.3} / 1 \\
DeepSeek-r1    & \underline{609} & \underline{622.5} & \textbf{642} & 5400.8 & \underline{624.5} / 2 \\
DeepSeek-v3    & 293.5 & 275.5 & 202.5 & 2491.2 & 257.2 / 3 \\
GPT-4o         & 232.5 & 245 & 223.5 & 2024.5 & 233.7 / 4 \\
Llama-3.1-405b & 16  & 20   & 115 & 1741.1 & 50.3 / 5 \\
\bottomrule
\end{tabular}
}
}
\end{table}

\subsection{Error Analysis}

\subsubsection{Impact of Task Size}

\begin{figure}
  \centering
  \begin{subfigure}[b]{0.49\textwidth}
    \centering
    \includegraphics[width=\textwidth]{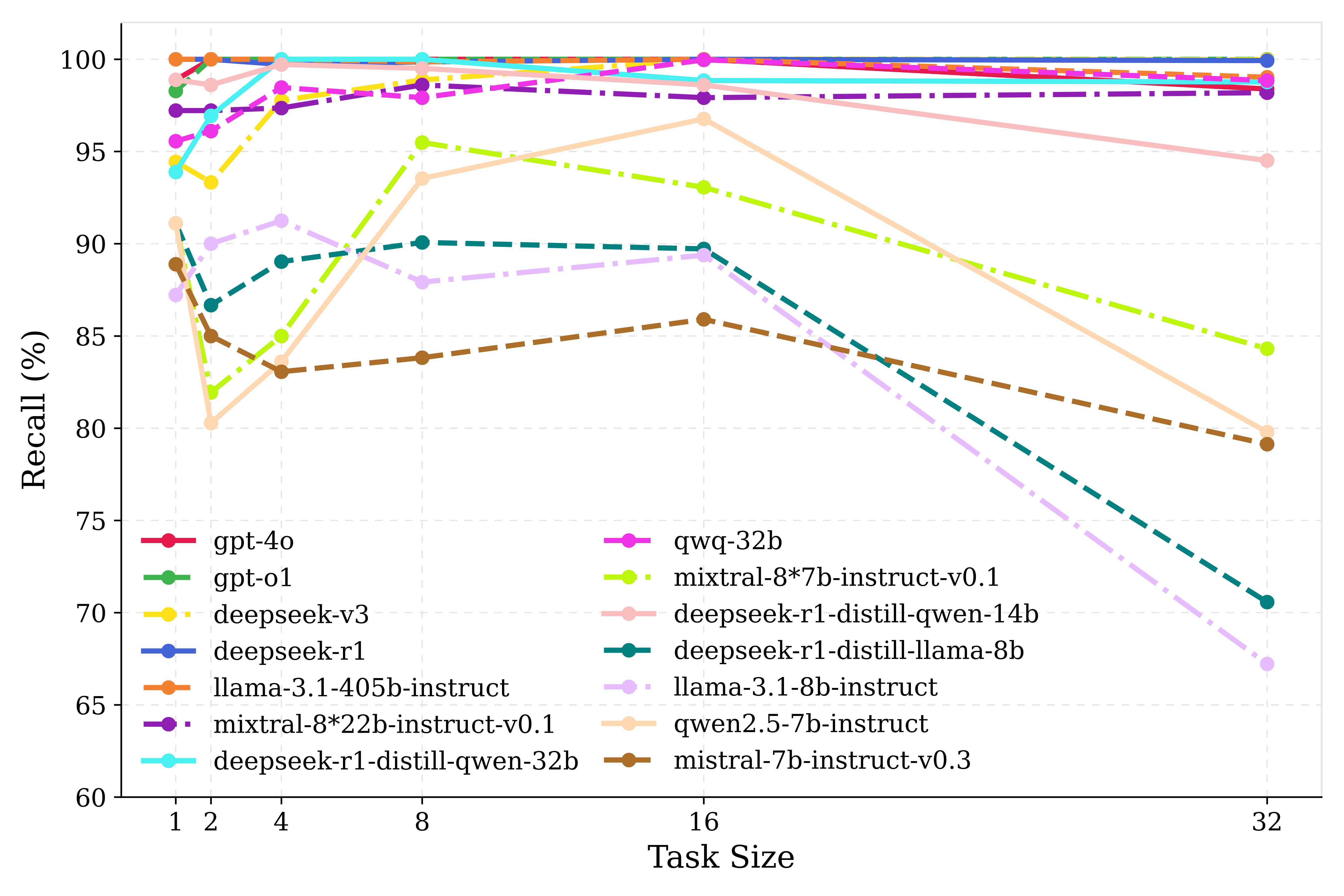}
    \caption{Information extraction task.}
    \label{fig:tasksize1}
  \end{subfigure}
  \hfill
  \begin{subfigure}[b]{0.49\textwidth}
    \centering
    \includegraphics[width=\textwidth]{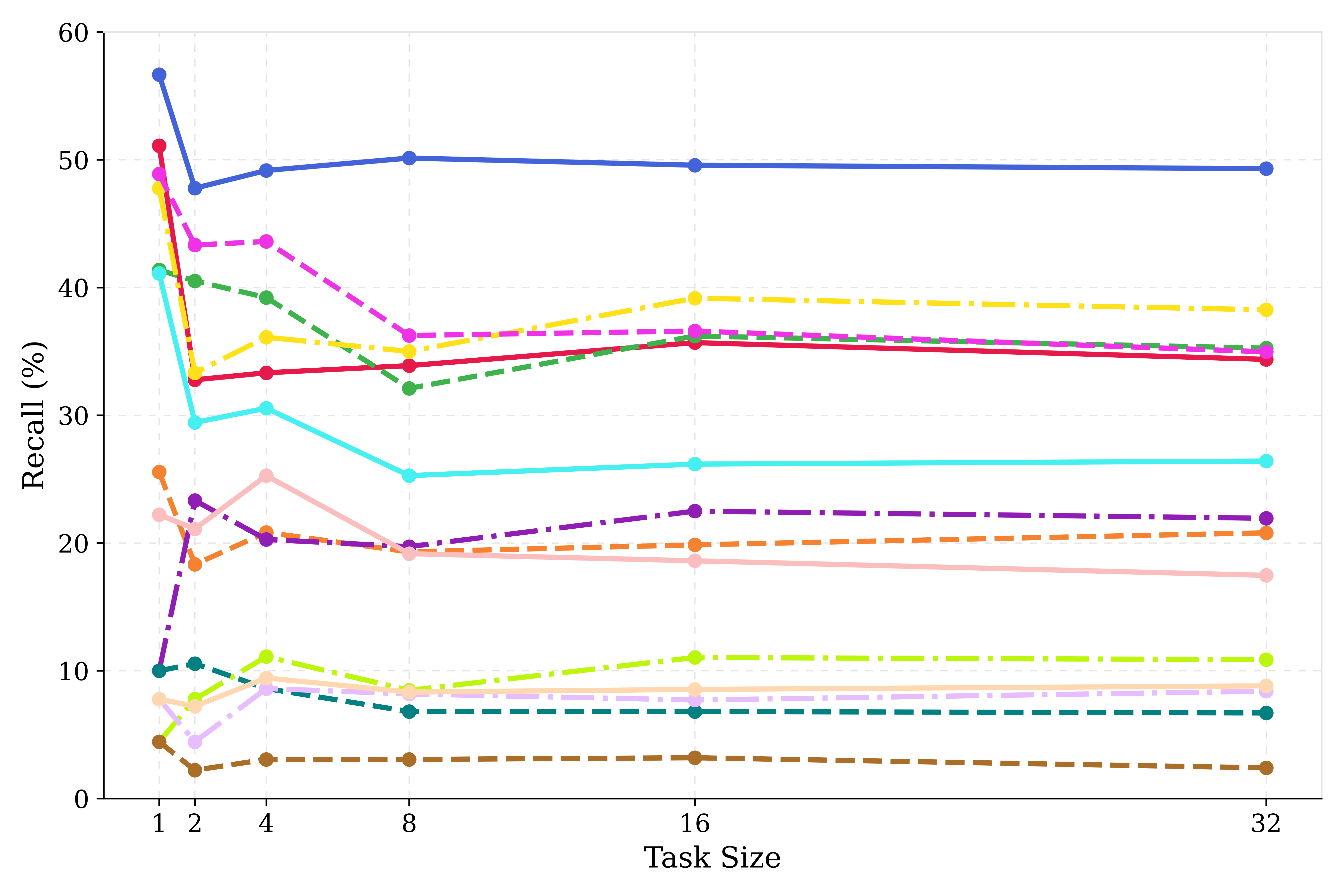}
    \caption{Indicator computation task.}
    \label{fig:tasksize2}
  \end{subfigure}
  \caption{Task size v.s. recall.}
  \label{fig:tasksize}
\end{figure}

In the information extraction and indicator computation tasks, we initially designed the prompt to request 32 random financial items at once. LLMs might struggle to maintain accuracy when handling too many items simultaneously. To verify this, we vary the number of items requested per prompt (we term it task size) across \{1, 2, 4, 8, 16, 32\}, and measure the recall under each setting. 
In Figure~\ref{fig:tasksize}, we find that recall generally decreases as task size increases, though the trend is not strictly monotonic. In the fact extraction task, large-sized LLMs maintain over 95\% recall even at size 32, showing strong robustness. In contrast, small-sized LLMs suffer 10 to 20 percentage point drops, indicating challenges in processing multiple items in one prompt. Indicator computation is more sensitive to task size. Even top-performing models show 5 to 10 point drops. Notably, small-sized LLMs show flatter trends—not due to higher robustness, but because their recall is already low, limiting the visible degradation. These findings suggest that large-sized LLMs handle multi-item prompts more effectively but still face error accumulation under load. Small-sized LLMs are more prone to omissions and miscalculations as task size grows.

\subsubsection{Effect of Numeric Tolerance}

Given LLMs inherently struggle with exact numeric calculations, we evaluate how  recall performance varies across different numerical error thresholds  $\theta$. Specifically, we test thresholds at \{0, 0.01, 0.05, 0.1, 0.2, 0.5, 0.8, 1.0\}. We observe that recall increases steadily as the threshold increases. The most dramatic improvement occurs between $\theta=0$ and $\theta=0.01$, indicating that many prediction errors are small and fall just outside of perfect accuracy. After $\theta=0.5$, the recall curve plateaus, showing diminishing returns. This suggests that most predictions fall within a 0–50\% relative error range, while values outside that are typically far off.

\begin{table}[ht]
\centering
\footnotesize
{
\caption{Recall under different numeric error tolerance thresholds for the indicator task.}
\label{tab: error}
\begin{tabular}{lcccccccc}
\toprule
\textbf{Model; $\theta$} & \textbf{0} & \textbf{0.01} & \textbf{0.05} & \textbf{0.1} & \textbf{0.2} & \textbf{0.5} & \textbf{0.8} & \textbf{1.0} \\
\midrule
GPT-4o & 34.38 & 63.69 & 75.18 & 80.90 & 86.67 & 90.71 & 91.41 & 91.47 \\
GPT-o1 & 35.26 & \textbf{79.76} & \textbf{90.56} & \textbf{92.63} & \textbf{94.30} & \textbf{95.74} & \textbf{96.05} & \textbf{96.08} \\
DeepSeek-v3 & \underline{38.26} & 62.10 & 72.48 & 77.95 & 83.50 & 88.42 & 89.37 & 89.50 \\
DeepSeek-r1 & \textbf{49.31} & \underline{76.43} & \underline{88.00} & \underline{90.45} & \underline{92.75} & \underline{94.50} & \underline{94.87} & \underline{94.91} \\
Llama-3.1-405b-instruct & 20.80 & 58.45 & 67.56 & 72.26 & 76.86 & 82.22 & 83.81 & 84.01 \\
Mixtral-8*22b-instruct-v0.1 & 21.94 & 46.45 & 53.81 & 57.04 & 60.72 & 66.51 & 69.64 & 70.48 \\
DeepSeek-r1-distill-qwen-32b & 26.42 & 62.03 & 73.62 & 79.51 & 85.06 & 88.55 & 89.24 & 89.38 \\
Qwq-32b & 34.97 & 67.46 & 79.76 & 84.81 & 89.33 & 92.28 & 92.83 & 92.87 \\
Mixtral-8*7b-instruct-v0.1 & 10.87 & 34.93 & 43.82 & 47.29 & 50.57 & 56.49 & 59.83 & 60.88 \\
DeepSeek-r1-distill-qwen-14b & 17.47 & 52.91 & 64.52 & 69.77 & 74.71 & 80.05 & 81.21 & 81.52 \\
DeepSeek-r1-distill-llama-8b & 6.70 & 29.34 & 34.05 & 35.56 & 37.76 & 43.13 & 46.01 & 47.07 \\
Llama-3.1-8b-instruct & 8.41 & 33.43 & 40.34 & 42.53 & 45.25 & 49.86 & 52.79 &  54.10 \\
Qwen2.5-7b-instruct & 8.82 & 35.81 & 46.59 & 50.45 & 54.01 & 60.06 &  62.88 & 63.82 \\
Mistral-7b-instruct-v0.3 & 2.40 & 9.38 & 18.93 & 22.06 & 24.99 & 29.62 & 31.94 & 32.53 \\
\bottomrule
\end{tabular}
}
\end{table}

\subsubsection{Effect of Knowledge Augmentation}

\begin{figure}
  \centering
  \begin{subfigure}[b]{0.49\textwidth}
    \centering
    \includegraphics[width=\textwidth]{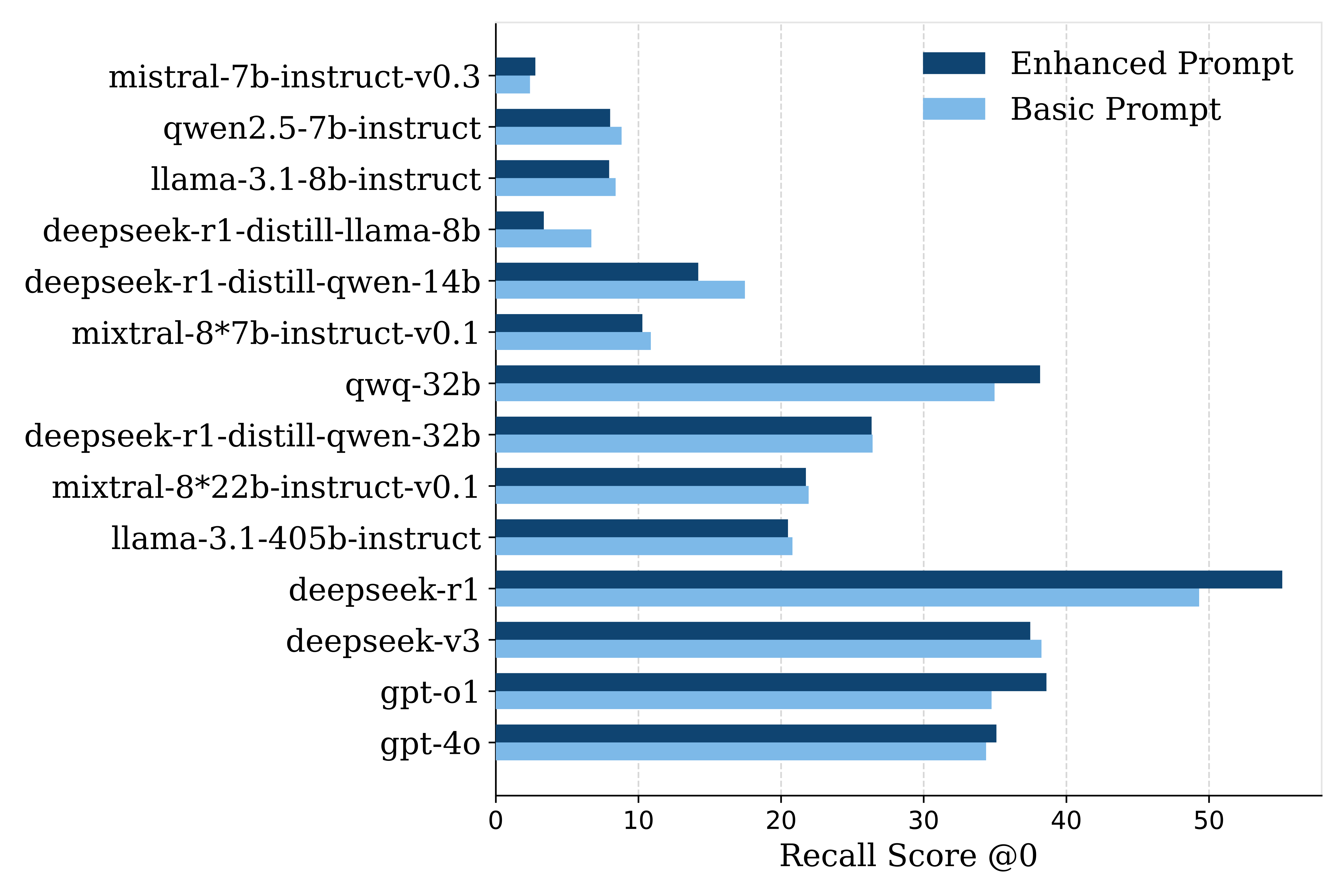}
    \caption{Error threshold $\theta$ = 0}
    \label{fig:ka_0}
  \end{subfigure}
  \hfill
  \begin{subfigure}[b]{0.49\textwidth}
    \centering
    \includegraphics[width=\textwidth]{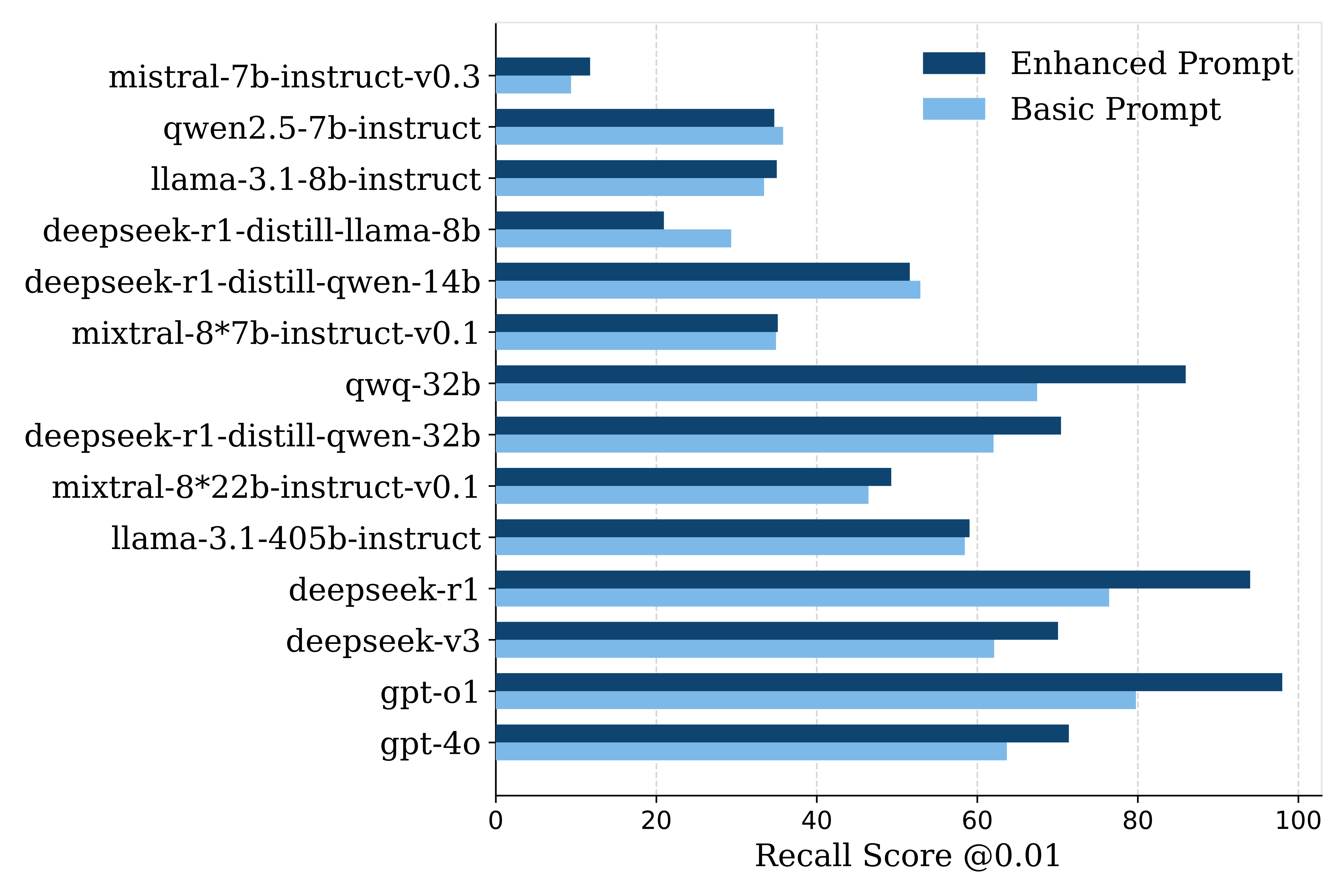}
    \caption{Error threshold $\theta$ = 0.01}
    \label{fig:ka_0.01}
  \end{subfigure}
  \caption{Performance comparison with and without knowledge augmentation.}
  \label{fig:knowledge}
\end{figure}

The financial knowledge encoded in LLMs may not align with the standard formulas we use to calculate financial indicators. We investigate whether explicitly providing calculation equations improves LLMs' performance on financial indicator tasks. In the "enhanced prompt" setting, we include exact calculation equations (e.g., "Net Profit Margin = Net Profit / Revenue"), while in the "basic prompt" setting, models are asked to compute indicators without supplementary information.
Figure \ref{fig:knowledge} illustrates the results of the evaluation under a strict numerical error threshold of $\theta=0$. Most medium-sized and large-sized models exhibit substantial improvements in recall when provided with enhanced prompts. 
In contrast, smaller models (e.g., Mistral-7b, DeepSeek-r1-distill-llama-8b) show minimal or even negative gains. This may stem from limited capacity in handling long structured inputs, where increased prompt complexity leads to confusion and degraded numerical reasoning. Notably, models with stronger reasoning abilities, such as DeepSeek-r1, GPT-o1, and Qwq-32b benefit the most from enhanced prompts. These models can better internalize explicit formula knowledge, leading to evidently improved performance.

\subsection{Case Study}
To evaluate the quality of financial reasoning produced by LLMs, we conduct a case study by manually reviewing outputs of GPT-o1 and DeepSeek-r1 in the logical reasoning task for a selected company. This evaluation is carried out by an financial expert, ensuring a comprehensive and credible assessment of the models' outputs. Reviews are summarized in Table~\ref{tab:llmjudge-case-study}. We find that current LLMs still show clear limitations in financial analysis. GPT-o1 has fewer logic errors than DeepSeek-r1 and both models perform poorly in terms of analytical depth and financial insight. This suggests that further optimization and domain-specific enhancement are needed for LLMs in financial applications.

\begin{table*}
\centering
\footnotesize
\caption{Case study of GPT-o1 and DeepSeek-r1 for a selected company.}
\label{tab:llmjudge-case-study}
\resizebox{\textwidth}{!}{
\begin{tabular}{>{\raggedright\arraybackslash}p{2.5cm} >{\raggedright\arraybackslash}p{5cm} >{\raggedright\arraybackslash}p{7cm}}
\toprule
\textbf{Dimension} & \textbf{GPT-o1} & \textbf{DeepSeek-r1} \\
\midrule
\textbf{Consistency} & 
- Interpretations are basically correct but lacks concrete numerical references. \newline
- Partial interpretation: "net profit $<$ operating cash flow" implies "strong cash collection capability". It ignores other influential factors of cash collection capability.
&
- Logic contradiction: A small margin increase (from 7.98\% to 10.98\%) is interpreted as "significant deterioration". \newline
- Misinterpretation: fixed assets rising from 20.21\% to 20.63\% is interpreted as large capital expenditure.
\newline
- Misunderstanding of terminology: "..., indicating a diminished capacity of cash flow to cover profits". Cash flow does not need to cover profits. It only needs to cover loan repayments and interest. \\
\midrule
\textbf{Depth of Analysis} &
- Analysis is plain and obvious but avoids major logic flaws. \newline  
- Indicators are vaguely interpreted. For instance, return on equity decline is mentioned without a deeper analysis, such as a DuPont breakdown. 
&
- Tends to force causal links between unrelated indicators and cause the incorrect analysis: "Inventory decreased 19.5\% but still higher than fixed assets" is interpreted as inventory obsolescence risk. In fact, inventory decrease is more likely due to sales improvement.\newline
- Incorrect causal inference: "Net financing cash flow was -1.28B due to debt repayment (245.92B) exceeding new borrowings (246.56B) " leading to the conclusion of "active deleveraging". The simultaneous occurrence of repayments and borrowings, offsetting one another, seems to reflect refinancing, not deleveraging.
\\
\midrule
\textbf{Financial Insight} & 
- Generally reasonable but lacks informative insight and diagnosis. &

- Insights based on flawed logic often compound prior errors. \\
\midrule
\textbf{Conclusion}
& - Basic financial understanding.
& - No practical use value.
 \\
\bottomrule
\end{tabular}
}
\end{table*}

\section{Limitation}
\label{sec: limitation}
Despite the valuable contributions of this study, we acknowledge the following limitations:
\begin{itemize}

    \item \textbf{LLMs limitation:} We have accurately extracted the necessary data for LLMs. In practice, financial analysts must find and navigate through a company's annual report, which usually contains over 200 pages. While this far exceeds typical LLM context limits, we will explore the possibility of benchmarking  LLM-based agents in a fully automated workflow.

    \item \textbf{Impicit evaluation:} In order to provide a rigorous evaluation, this study investigates the capability of LLMs for conducting financial statement analysis implicitly, where we instruct LLMs to generate intermediate results throughout the financial statement analysis process. 
    \item \textbf{Resource constraints:} Due to budget constraints, this study evaluates limited proprietary LLMs including GPT-4o and GPT-o1. We will hold a benchmark leaderboard, and invite more proprietary LLMs to participate.
\end{itemize}
\section{Conclusion}
In this work, we present \benname{}, a task-oriented benchmark dataset for evaluating LLMs in financial fundamental analysis, with the first set of tasks being the financial statement analysis. We evaluate 14 different LLMs on information extraction, indicator computation, and logic reasoning tasks. Although these key tasks involved in financial statement analysis are relatively intuitive for humans, current LLMs demonstrate innate limitations in performing these tasks, particularly in satisfying the domain’s strict requirements for exceptional precision and the complete intolerance of hallucinations. Given the specific characteristics of applying LLMs in finance,  we will continue our work to cover more fundamental analysis competence and extend to the evaluation of the financial task capabilities of LLM-based agents in the future.

\section*{Acknowledgment}
We are thankful to OpenBayes.com for generously providing the computational resources and support that made our experiments possible.

\bibliographystyle{unsrt}

\clearpage

\appendix
\begin{center}
\Large \textbf{Appendix}
\end{center}

\section{Descriptive Statistics of FinAR-Bench}

FinAR-Bench consists of a total of 1,170 financial analysis tasks from 100 companies, divided into a development set with 90 companies and a  test set with 10 companies. Each company contributes 13 tasks: 6 factual extraction tasks, 6 indicator computation tasks, and 1 reasoning task.
Factual and indicator tasks are defined over six task sizes (1, 2, 4, 8, 16, 32 tables), enabling evaluation of LLM performance under different task sizes. The reasoning task requires logical interpretation over given conditions. All tasks are constructed from real-world annual reports covering fiscal years 2022 and 2023. The development set is publicly released, while the test set is reserved for benchmarking and leaderboard evaluations.

\begin{table}[ht]
\centering
\footnotesize
\caption{Summary Statistics of FinAR-Bench}
\label{tab:finar_statistics}
\begin{tabular}{l r}
\toprule
\textbf{Metric} & \textbf{Value} \\
\midrule
Total Companies & 100 \\
Dev / Test Split & 90 / 10 \\
Tasks per Company & 13 \\
Total Tasks & 1,170 \\
Fact Tasks (per company) & 6 \\
Indicator Tasks (per company) & 6 \\
Reasoning Tasks (per company) & 1 \\
Fiscal Years Covered & 2022, 2023 \\
\bottomrule
\end{tabular}
\end{table}

\section{Task Details}

\label{app:benchmark-tasks}
This appendix summarizes the financial items, indicators, and reasoning conditions used in the FinAR-Bench evaluation. It consists of three components:

\subsection*{Financial Items}
Table~\ref{tab:fact_financial_items} lists the financial items involved in all tasks.

\FloatBarrier
\begin{table*}[ht]
\centering
\caption{Financial Items}
\resizebox{0.98\textwidth}{!}{
\footnotesize
\begin{tabular}{p{0.03\columnwidth}p{0.42\columnwidth}p{0.03\columnwidth}p{0.42\columnwidth}}
\toprule
\textbf{No.} & \textbf{Item} & \textbf{No.} & \textbf{Item} \\
\midrule
1  & Total Revenue & 17 & Net Value of Fixed Assets \\
2  & Total Cost of Goods Sold & 18 & Construction in Progress \\
3  & Revenue & 19 & Short-term Loans \\
4  & Operating Cost & 20 & Long-term Debt \\
5  & Total Profit & 21 & Bonds Payable \\
6  & Total Assets & 22 & Net Income \\
7  & Total Liabilities & 23 & Net Income Attributable to Parent \\
8  & Total Current Assets & 24 & Selling Expenses \\
9  & Total Non-Current Assets & 25 & Administrative Expenses \\
10 & Total Current Liabilities & 26 & Financial Expenses \\
11 & Total Long-Term Liabilities & 27 & Accounts Receivable \\
12 & Total Shareholders' Equity Including Minority Interest & 28 & Accounts Payable \\
13 & Total Shareholders' Equity Excluding Minority Interest & 29 & Payroll Payable \\
14 & Money Capital & 30 & Net Cash Provided by Operating Activities \\
15 & Goodwill & 31 & Net Cash Used in Investing Activities \\
16 & Inventories & 32 & Net Cash Provided Used in Financing Activities \\
\bottomrule
\end{tabular}
}
\label{tab:fact_financial_items}
\end{table*}

\subsection*{Financial Indicators and Calculation Formulas}
Table~\ref{tab:financial_indicators} summarizes the financial indicators used in our evaluation tasks, along with their corresponding calculation formulas.

\FloatBarrier
\begin{table*}[htbp]
\centering
\caption{Financial Indicators and Calculation Formulas}
\resizebox{0.98\textwidth}{!}{
\footnotesize
\begin{tabular}{@{}l p{5cm} p{7.8cm}}
\toprule
\textbf{No.} & \textbf{Indicator} & \textbf{Formula} \\
\midrule
1 & Return on Equity & Net Income Attributable to Parent / Average Equity Attributable to Parent \\
2 & Return on Assets & Net Income / Average Total Assets \\
3 & Gross Profit Margin & (Operating Revenue - Operating Cost) / Operating Revenue \\
4 & Net Profit Margin & Net Income / Operating Revenue \\
5 & Debt to Assets Ratio & Total Liabilities / Total Assets \\
6 & Current Ratio & Current Assets / Current Liabilities \\
7 & Quick Ratio & (Current Assets - Inventories - Prepayments - Current Portion of Non-current Assets - Other Current Assets) / Current Liabilities \\
8 & Period Expense Ratio & Gross Profit Margin - Net Profit Margin \\
9 & Equity Multiplier & Total Assets / Total Shareholders' Equity \\
10 & Equity Ratio & Total Liabilities / Equity Attributable to Parent \\
11 & Inventory Turnover Days & 360 / Inventory Turnover Ratio \\
12 & Accounts Receivable Turnover Days & 360 / Accounts Receivable Turnover Ratio \\
13 & Accounts Payable Turnover Days & 360 / Accounts Payable Turnover Ratio \\
14 & Operating Cycle & Inventory Turnover Days + Accounts Receivable Turnover Days \\
15 & Total Asset Turnover Ratio & Operating Revenue / Average Total Assets \\
16 & Inventory Turnover Ratio & Operating Cost / Average Inventories \\
17 & Accounts Receivable Turnover Ratio & Operating Revenue / Average Accounts Receivable \\
18 & Accounts Payable Turnover Ratio & (Ending Inventories + Operating Cost - Beginning Inventories) / Average Accounts Payable \\
19 & Current Asset Turnover Ratio & Operating Revenue / Average Current Assets \\
20 & Fixed Asset Turnover Ratio & Operating Revenue / Average Net Fixed Assets \\
21 & Net Profit Growth Rate & (Current Year Net Income - Previous Year Net Income) / Previous Year Net Income \\
22 & Operating Revenue Growth Rate & (Current Year Revenue - Previous Year Revenue) / Previous Year Revenue \\
23 & Net Operating Cash Flow Growth Rate & (Current Year Operating Cash Flow - Last Year Operating Cash Flow) / Last Year Operating Cash Flow \\
24 & Accounts Receivable Growth Rate & (Ending Accounts Receivable - Beginning Accounts Receivable) / Beginning Accounts Receivable \\
25 & Net Cash Flow to Net Profit Ratio & Net Operating Cash Flow / Net Income \\
26 & Operating Cash Inflow to Revenue Ratio & Cash Received from Sale of Goods and Services / Operating Revenue \\
27 & Selling Expenses to Revenue Ratio & Selling Expenses / Operating Revenue \\
28 & Administrative Expenses to Revenue Ratio & Administrative Expenses / Operating Revenue \\
29 & Financial Expenses to Revenue Ratio & Financial Expenses / Operating Revenue \\
30 & Goodwill to Total Assets Ratio & Goodwill / Total Assets \\
31 & Accounts Receivable to Total Assets Ratio & Accounts Receivable / Total Assets \\
32 & Accounts Payable to Total Liabilities Ratio & Accounts Payable / Total Liabilities \\
33 & Inventory to Total Assets Ratio & Inventories / Total Assets \\
34 & Fixed Assets to Total Assets Ratio & Net Fixed Assets / Total Assets \\
35 & Current Assets to Total Assets Ratio & Current Assets / Total Assets \\
36 & Current Liabilities to Total Liabilities Ratio & Current Liabilities / Total Liabilities \\
37 & Non-current Assets to Total Assets Ratio & Non-current Assets / Total Assets \\
38 & Non-current Liabilities to Total Liabilities Ratio & Non-current Liabilities / Total Liabilities \\
\bottomrule
\end{tabular}
}
\label{tab:financial_indicators}
\end{table*}

\subsection*{Reasoning Task Conditions}
Table~\ref{tab:reasoning_tasks_conditions} lists the reasoning task conditions in our benchmark. These tasks require models to interpret trends, compare values, or apply threshold-based logic to financial indicators across different fiscal years.

\FloatBarrier
\begin{table*}[ht]
\centering
\caption{Reasoning Task Conditions}
\resizebox{0.98\textwidth}{!}{
\footnotesize
\begin{tabular}{@{}p{1cm}p{11cm}@{}}
\toprule
\multicolumn{2}{c}{\textbf{Trend-Based Judgments}} \\
\midrule
1/2 & Return on Equity (ROE) is increasing / decreasing \\
3/4 & Return on Assets (ROA) is increasing / decreasing \\
5/6 & Net Profit Margin is increasing / decreasing \\
7/8 & Gross Profit Margin is increasing / decreasing \\
9/10 & Operating Expense Ratio is increasing / decreasing \\
11/12 & Equity Multiplier is increasing / decreasing \\
13/14 & Debt to Assets Ratio is increasing / decreasing \\
15/16 & Current Ratio is increasing / decreasing \\
17/18 & Quick Ratio is increasing / decreasing \\
19/20 & Administrative Expenses to Revenue Ratio is increasing / decreasing \\
21/22 & Selling Expenses to Revenue Ratio is increasing / decreasing \\
23/24 & Financial Expenses to Revenue Ratio is increasing / decreasing \\
25/26 & Current Assets to Total Assets Ratio is increasing / decreasing \\
27/28 & Non-current Assets to Total Assets Ratio is increasing / decreasing \\
29/30 & Current Liabilities to Total Liabilities Ratio is increasing / decreasing \\
31/32 & Long-term Liabilities to Total Liabilities Ratio is increasing / decreasing \\
33/34 & Goodwill to Total Assets Ratio is increasing / decreasing \\
35/36 & Accounts Receivable to Total Assets Ratio is increasing / decreasing \\
37/38 & Accounts Payable to Total Liabilities Ratio is increasing / decreasing \\
39/40 & Net Fixed Assets to Total Assets Ratio is increasing / decreasing \\
\midrule
\multicolumn{2}{c}{\textbf{Threshold-Based Judgments}} \\
\midrule
41/42 & Operating Revenue Growth Rate $>$ 0 / $<$ 0 \\
43/44 & Net Profit Growth Rate $>$ 0 / $<$ 0 \\
45/46 & Operating Cash Flow Growth Rate $>$ 0 / $<$ 0 \\
47/48 & Accounts Receivable Growth Rate $>$ 0 / $<$ 0 \\
49/50 & Net Operating Cash Flow $>$ 0 / $<$ 0 \\
51/52 & Net Investing Cash Flow $>$ 0 / $<$ 0 \\
53/54 & Net Financing Cash Flow $>$ 0 / $<$ 0 \\
\midrule
\multicolumn{2}{c}{\textbf{Comparative Judgments}} \\
\midrule
55/56 & Operating Revenue Growth Rate $>$ / $<$ Net Profit Growth Rate \\
57/58 & Net Profit Growth Rate $>$ / $<$ Operating Cash Flow Growth Rate \\
59/60 & Inventories $>$ / $<$ Net Fixed Assets \\
61/62 & Net Profit $>$ / $<$ Net Operating Cash Flow \\
63/64 & Selling Expenses $>$ / $<$ Administrative Expenses \\
\bottomrule
\end{tabular}
}
\label{tab:reasoning_tasks_conditions}
\end{table*} 

\section{Performance of Large Models on PDF Format}

In addition to the main experiments on clean text-format inputs, we evaluate the robustness of large-scale LLMs (sizes \verb|>|100B) on PDF-format annual reports, which more closely resemble real-world usage scenarios. Small and medium-sized models are excluded due to the model's context length limitation or its inability to generate long content.

Table~\ref{tab:pdf-evaluation} presents the precision and recall of six large models across the three task types. While information extraction remains relatively strong (e.g., DeepSeek-r1 reaches 96.44\% precision), indicator computation shows significant degradation, with most models achieving under 35\% recall. Logic reasoning tasks remain highly variable, with GPT-o1 reaching 85.73\% recall, while Mixtral-8x22B drops below 55\%.

These results highlight the difficulty of parsing structured content directly from PDFs and the performance bottlenecks in numerical computation and reasoning when layout noise is introduced. Improving layout-aware modeling and financial pretraining remains a key direction for future work.

\begin{table*}[htbp]
\centering
\caption{Precision and recall of the largest models (Size=L) under PDF input.}
\label{tab:pdf-evaluation}
\resizebox{0.98\textwidth}{!}{
\footnotesize
\begin{tabular}{ll
cc  
cc  
cc  
}
\toprule
\textbf{Size} & \textbf{Model} 
& \multicolumn{2}{c}{\textbf{Information Extraction}} 
& \multicolumn{2}{c}{\textbf{Indicator Computation}} 
& \multicolumn{2}{c}{\textbf{Logic Reasoning}} \\
\cmidrule(lr){3-4} \cmidrule(lr){5-6} \cmidrule(lr){7-8}
& 
& P & R 
& P & R 
& P & R \\
\midrule
\textbf{L} & GPT-4o & 94.90 & 93.89 & 33.78 & 33.78 & 60.61 & 60.26 \\
& GPT-o1 & \underline{96.08} & \underline{96.08} & 33.78 & 33.78 & \textbf{85.73} & \textbf{85.73} \\
& DeepSeek-v3 & 93.42 & 93.42 & \underline{35.28} & \underline{35.28} & 58.35 & 58.72 \\
& DeepSeek-r1 & \textbf{96.44} & \textbf{96.20} & \textbf{48.09} & \textbf{47.63} & \underline{84.40} & \underline{73.40} \\
& Llama-3.1-405b-instruct & 95.67 & 92.61 & 18.65 & 18.63 & 61.78 & 55.40 \\
& Mixtral-8*22b-instruct-v0.1 & 85.82 & 85.03 & 12.24 & 12.22 & 54.25 & 54.25 \\
\bottomrule
\end{tabular}
}
\end{table*}

\section{PDF Extractor Results}
\label{sec: pdf}
Table \ref{tab: pdf} reports the performance of different PDF extraction methods for financial statement data. The evaluation is based on the fact extraction task. We assess the performance of six different PDF extraction methods: pdfplumber, pdfminer, pypdf, pdftotext, minerU, and pymupdf. The results are shown in terms of precision and recall for each method. 

\begin{table*}[ht]
\centering
\caption{PDF Extractor Results: Precision and Recall for Fact Extraction.}
\label{tab: pdf}
\resizebox{0.98\textwidth}{!}{
\footnotesize
\begin{tabular}{p{1.5cm}p{1.1cm}p{1cm}p{1.1cm}p{1cm}p{1.1cm}p{1cm}p{1.1cm}p{1cm}}
\toprule
\textbf{Method} & 
\multicolumn{2}{c}{\textbf{DeepSeek-r1}} & 
\multicolumn{2}{c}{\textbf{DeepSeek-v3}} & 
\multicolumn{2}{c}{\makecell{\textbf{Llama-3.1-}\\\textbf{405b-instruct}}} & 
\multicolumn{2}{c}{\makecell{\textbf{Mixtral-8x22b-}\\\textbf{instruct-v0.1}}} \\
\cmidrule(r){2-3} \cmidrule(r){4-5} \cmidrule(r){6-7} \cmidrule(r){8-9}
 & Precision & Recall & Precision & Recall & Precision & Recall & Precision & Recall\\
\midrule
pdfplumber & 95.52 & 95.19 & 95.89 & 95.53 & 95.89 & \textbf{95.53} & 85.24 &  \textbf{85.45}  \\
pdfminer   & 66.19 & 66.03 & 52.39 & 52.39 & 59.98 & 57.36 & 7.59 & 7.41  \\
pypdf      & \underline{96.10}  & \underline{95.77} & \underline{95.86} & \underline{95.86} & \underline{96.37} & 91.75 & 84.34  & 84.75 \\
pdftotext  & 76.79 & 76.62 & 63.56 & 63.56 & 32.08 & 31.83 & 8.70 & 7.62 \\
minerU     & 94.64 & 94.45 & 93.11 & 93.11 & 93.49 & 88.91 & 77.05 & 75.50 \\
pymupdf    & \textbf{96.45}  & \textbf{96.22} & \textbf{95.98} & \textbf{95.98} & \textbf{96.55} & \underline{93.48} & \textbf{85.83} & \underline{85.05} \\
\bottomrule
\end{tabular}
}
\end{table*}

\end{document}